\documentclass[aps,prd,twocolumn]{revtex4-2}
\usepackage{amsfonts}
\usepackage{amssymb}
\usepackage{amsmath}
\usepackage{color}
\usepackage[usenames,dvipsnames]{xcolor}
\usepackage{float}
\usepackage{accents}
\usepackage{graphicx}
\usepackage{epstopdf}
\usepackage{soul}
\usepackage[colorlinks=true,pdfstartview=FitV,linkcolor=blue,citecolor=blue,urlcolor=blue,breaklinks=true]{hyperref}

\begin{document}

\title{BPS chiral vortices in a Maxwell-Higgs electrodynamics}
\author{J. Andrade}
\email{joao.luis@discente.ufma.com.br}
\email{joao.luis@ufma.br}
\author{Rodolfo Casana}
\email{rodolfo.casana@ufma.br}
\email{rodolfo.casana@gmail.com}
\affiliation{Departamento de F\'{\i}sica, Universidade Federal do Maranh\~{a}o, {65080-805%
}, S\~{a}o Lu\'{\i}s, Maranh\~{a}o, Brazil.}
\author{E. da Hora}
\email{carlos.hora@ufma.br}
\email{edahora.ufma@gmail.com }
\affiliation{Coordenadoria Interdisciplinar de Ci\^{e}ncia e Tecnologia, Universidade
Federal do Maranh\~{a}o, {65080-805}, S\~{a}o Lu\'{\i}s, Maranh\~{a}o,
Brazil.}

\begin{abstract}
We investigate the existence of BPS structures in a Maxwell-Higgs electrodynamics immersed within a chiral medium, whose electromagnetic properties are described by both a the Chern-Simons term and a neutral scalar field. The implementation of the Bogomol'nyi-Prasad-Sommerfield's technique provides the BPS potential and the self-dual equations whose solutions saturate the Bogomol'nyi bound. In such a context, we look for vortices in two chiral media: the first  one engenders localized vortices with an
exponential decay similar to that of the Abrikosov-Nielsen-Olesen  solutions, whereas the second medium generates delocalized  profiles whose tail follows a power-law decay.  Once we have solved the BPS systems, we comment on the
effects induced by the presence of the chiral medium on the Maxwell-Higgs
vortices.
\end{abstract}

\maketitle

\section{Introduction}

\label{Intro}

In the context of classical field theories, configurations with nontrivial
topology are solutions of highly nonlinear second-order field equations
driven by the presence of a symmetry-breaking potential which characterizes
a phase transition \cite{n5}. However, under exceptional circumstances, they
can also be studied as the solutions of a set of coupled first-order
differential equations obtained from the minimization of the effective
energy functional through the Bogomol'nyi-Prasad-Sommerfield (BPS) formalism
\cite{n4}. Other methods which also lead to the first-order equations
include the study of the conservation of the energy-momentum tensor \cite{ano} and the so-called On-Shell Method \cite{onshell}. In this scenario, topologically nontrivial configurations in a planar space (so-called
\textit{vortices}) were obtained not only in the Maxwell-Higgs (MH) model
\cite{n1}, but also in the Chern-Simons-Higgs (CSH) \cite{cshv} and in the
Maxwell-Chern-Simons-Higgs (MCSH) \cite{mcshv} ones.

Another interesting issue is the existence of first-order vortices in the
context of enlarged gauged models. In this case, it is worthwhile
to highlight that the usual Maxwell-Higgs model itself was extended to
accommodate an $SO(3)$ group  via the inclusion of an extra scalar
sector \cite{witten}. In addition, a recent work \cite{n44} has verified
that such enlarged theory indeed supports first-order vortices with internal structures, which may find relevant applications in connection with the study of metamaterials \cite{n45}. Beyond this point, such vortices  were also shown to occur in an extended Maxwell-$CP(2)$ model \cite{joao}.

The present manuscript aims to investigate the existence of BPS configurations within a chiral electrodynamics defined by a generalized Maxwell-Chern-Simons model, the chiral one prototype. For our purpose, we have organized our results as follows: In the Sec. II, we define a chiral electrodynamics coupled to the Higgs field and a real scalar field in the context of an extended Maxwell-Chern-Simons-Higgs theory. We then focus our attention on the construction of the BPS structure inherent to the resulting chiral electrodynamics, from which we find the self-dual equations and the respective Bogomol'nyi bound for the total energy. Next, in the Sec. III, the system formed by the BPS equations and the Gauss law are solved through a finite-difference technique to obtain chiral vortices for two different choices of  a dielectric function. We depict the relevant field profiles and identify the effects of the chiral medium. In the Section IV, we end our manuscript by summarizing our results and enunciating our perspectives regarding future
investigations.

%%%%%%%%%%%%%%%%%%%%%%%%

\section{The Maxwell-Higgs electrodynamics in a chiral medium\label{2}}

It is well-known that in $(1+3)$-dimensions, the following Lagrangian
density describes the electromagnetic field in the vacuum \cite{foot1}
\begin{equation}
\mathcal{L}_{\text{v}}=-\frac{1}{4}F_{\hat{\mu}\hat{\nu}}F^{\hat{\mu}\hat{\nu}} \text{,}  \label{eqch1}
\end{equation}%
while, in an electromagnetic medium, the constitutive relations
between the electric and magnetic fields are expressed by the tensor $\mathcal{C}_{%
\hat{\mu}\hat{\nu}}$, i.e.,
\begin{equation}
\mathcal{C}_{\hat{\mu}\hat{\nu}}\equiv \mathcal{C}_{\hat{\mu}\hat{\nu}}\left( A_{\hat{\mu}},F_{%
\hat{\mu}\hat{\nu}}\right) \text{,}  \label{eqch2}
\end{equation}%
where $A_{\hat{\mu}}$ is the electromagnetic field and $F_{\hat{\mu}\hat{\nu}} =\partial _{\hat{\mu}}A_{\hat{\nu}}-\partial _{\hat{\nu}}A_{\hat{\mu}}$
represents its strength tensor. Then, the Lagrangian density which
describes the electromagnetic field in the medium in the presence of a
source becomes
\begin{equation}
\mathcal{L}_{\text{m}}=-\frac{1}{4}\mathcal{C}_{\hat{\mu}\hat{\nu}}F^{\hat{\mu}\hat{\nu%
}} +A_{\hat{\mu}}J^{\hat{\mu}}\text{.}  \label{eqch3}
\end{equation}%

A constitutive tensor $\mathcal{C}_{\hat{\mu}\hat{\nu}}$ engendering a chiral medium
\cite{qed_chiral} is given by
\begin{equation}
\mathcal{C}_{\mu \nu }=F_{\hat{\mu}\hat{\nu}}+\frac{1}{4}\epsilon _{\hat{\mu}\hat{\nu}{%
\hat{\rho}}{\hat{\sigma}}}b^{\hat{\rho}}A^{\hat{\sigma}}\text{,}
\label{eqch5}
\end{equation}
from which one gets that a resulting Lagrangian density describing a chiral electrodynamics reads
\begin{equation}
\mathcal{L}_{\text{m}}=-\frac{1}{4}F_{\hat{\mu}\hat{\nu}}F^{\hat{\mu}\hat{\nu}} -\frac{1}{4}\epsilon _{\hat{\mu}\hat{\nu}{\hat{\rho}}{\hat{\sigma}}} b^{\hat{\rho}}A^{\hat{\sigma}}F^{\hat{\mu}\hat{\nu}}+ A_{\hat{\mu}} J^{\hat{\mu}}\text{,}  \label{eqch5a}
\end{equation}%
where $\epsilon ^{\hat{\mu}\hat{\nu}{\hat{\rho}}{\hat{\sigma}}}$ is the
Levi-Civita tensor (with $\epsilon ^{0123}=+1$) and $b^{\hat{\mu}}=\left(
b_{0},\mathbf{b}\right) $ represents a vector whose component $b_{0}$ is
related to the chiral magnetic effect \cite{CQED1,CQED2}, while the vector $\mathbf{b}$ is associated to the anomalous Hall effect \cite{CQED3, CQED4,
CQED5, CQED6, CQED7} or also can be related to the generation of an anomalous charge density \cite{Wilzeck}.  In the context of Lorentz-violating field theories, the vector $b^{\hat{\mu}}$ becomes the Carroll-Field-Jackiw background $\left( k_{AF}\right)^{\hat{\mu}}$, from which the corresponding model is named the Maxwell-Carroll-Field-Jackiw electrodynamics (MCFJ) \cite{CQED8}. Furthermore, the axion electrodynamics is also described by the Lagrangian density (\ref{eqch3}) when the field $b_{\hat{\mu}}= \partial_{\hat{\mu}}\theta $, where $\theta $ is a pseudoscalar function identified as the axion field \cite{axion1,axion2}. Recently, the MCFJ electrodynamics was devoted to analyzing the effects of the chiral medium on Cerenkov's radiation \cite{Urrutia} and the cold plasma modes \cite{Filipe1, Filipe2}.

We now consider the electromagnetic field in a chiral medium interacting with the Higgs field $\phi $ and a real scalar field $\chi$ through the Lagrangian density
\begin{eqnarray}
\mathcal{L} &=&-\frac{G(\chi )}{4}F_{\hat{\mu}\hat{\nu}}F^{\hat{\mu}\hat{\nu}%
}-\frac{1}{4}\epsilon _{\hat{\mu}\hat{\nu}{\hat{\rho}}{\hat{\sigma}}}b^{\hat{%
\rho}}A^{\hat{\sigma}}F^{\hat{\mu}\hat{\nu}}  \notag \\
&&+\left\vert D_{\hat{\mu}}\phi \right\vert ^{2}+\frac{1}{2}\partial _{\hat{%
\mu}}\chi \partial ^{\hat{\mu}}\chi -V\left( \left\vert \phi \right\vert
,\chi \right) \text{,}  \label{eqch7x}
\end{eqnarray}
where $G(\chi)$ plays the role of a dielectric function, whereas $D_{\hat\mu}\phi$ is the minimal covariant derivative which couples the electromagnetic and Higgs fields, i.e.,
\begin{equation}
D_{\hat{\mu}}\phi =\partial _{\hat{\mu}}\phi -ieA_{\hat{\mu}}\phi \text{,}
\label{eqch6}
\end{equation}
with $e$ being the electromagnetic coupling constant.

Here, we point out that when $G(\chi)\equiv 1$, the field $\chi$ decouples of the model, therefore originating the chiral Maxwell-Higgs electrodynamics or, in a Lorentz-violating context, the Maxwell-Carroll-Field-Jackiw-Higgs theory \cite{CQED9}. The Reference \cite{Alyson} has shown the existence of a BPS structure coming from a $(1+2)$-dimensional version of the Maxwell-Carroll-Field-Jackiw-Higgs model described by the Lagrangian density \cite{foot2}
\begin{eqnarray}
\mathcal{L} &=&-\frac{1}{4}F_{\mu \nu }F^{\mu \nu }-\frac{\kappa }{4}%
\,\epsilon ^{\nu \rho \sigma }A_{\nu }F_{\rho \sigma }+\left\vert D_{\mu
}\phi \right\vert ^{2}  \notag \\
&&+\frac{1}{2}\partial _{\mu }\Psi \partial ^{\mu }\Psi -e^{2}\Psi
^{2}\left\vert \phi \right\vert ^{2}-U\left( \left\vert \phi \right\vert
,\Psi \right)  \label{L2} \\
&&-\frac{1}{2}\epsilon ^{\mu \rho \sigma }\left( k_{AF}\right) _{\mu
}A_{\rho }\partial _{\sigma }\Psi -\frac{1}{2}\epsilon ^{\mu \rho \sigma
}\left( k_{AF}\right) _{\mu }\Psi ~\partial _{\rho }A_{\sigma },  \notag
\end{eqnarray}%
where $\kappa =-\left(k_{AF}\right)_{\hat{3}}$ plays the role of a Chern-Simons coupling constant,  $\epsilon ^{\mu\nu\rho}$ is the $(1+2)$-dimensional Levi-Civita tensor ($\epsilon ^{012}=+1$), $\Psi = A_{\hat{3}}$ {{stands for}} a real scalar field, and $U\left(\left\vert \phi \right\vert, \Psi\right) $ represents the potential. In such a context, $\left(k_{AF}\right) _{\mu}$ is the $(1+2)$-dimensional CFJ vector background which couples the real and gauge fields. We highlight that here $\Psi$ plays the role of the neutral field \cite{mcshv} introduced into the Maxwell-Chern-Simons-Higgs model to attain a BPS structure.

We aim to show the existence of a BPS structure (able to generate chiral vortices) starting from a $(1+2)$-dimensional version of the model (\ref{eqch7x}) attained  via the same dimensional reduction procedure used in the Ref. \cite{Alyson} and by considering $\left(k_{AF}\right)_{\mu}=0$ \cite{nextwork}. This way, we get the model described by the Lagrangian density
\begin{eqnarray}
\mathcal{L} &=&-\frac{G(\chi )}{4}F_{\mu \nu }F^{\mu \nu }-\frac{\kappa }{4}%
\epsilon ^{\mu \nu \kappa }A_{\mu }F_{\nu \kappa }+\left\vert D_{\mu }\phi
\right\vert ^{2}  \notag \\[0.2cm]
&&+\frac{G(\chi )}{2}\partial _{\mu }\Psi \partial ^{\mu }\Psi -e^{2}\Psi
^{2}\left\vert \phi \right\vert ^{2}  \notag \\[0.2cm]
&&+\frac{1}{2}\partial _{\mu }\chi \partial ^{\mu }\chi -U\left( \left\vert
\phi \right\vert ,\Psi ,\chi \right) \text{,}  \label{1}
\end{eqnarray}
which will be the starting point for our study. We immediately recognize it as representing an enlarged version of the self-dual MCSH scenario, where $U\left(\left\vert \phi \right\vert, \Psi, \chi \right) $ defines a potential to be determined by the BPS formalism itself. As we will below, the function $G(\chi)$, beyond specifying the electromagnetic properties of the medium, is also responsible for modifying the profiles of the electric and magnetic fields inherent to the vortex configurations. Also, we observe that when $G(\chi) =1$, the field $\chi$ decouples from the gauge sector and, as a consequence, the Lagrangian density (\ref{1}) reduces to that of the MCSH model in an electromagnetic vacuum. Moreover, we point out that the presence of the field $\chi$ has already been considered in the context of topological vortices with internal structures in both the Maxwell-Higgs and the Maxwell-$CP(2)$ models.

The Euler-Lagrange equation obtained from the Lagrangian density (\ref{1}) for  the gauge field reads
\begin{equation}
\partial _{\alpha }\left( GF^{\alpha \mu }\right) -\frac{\kappa }{2}\epsilon
^{\mu \alpha \beta }F_{\alpha \beta }=eJ^{\mu }\text{,}
\end{equation}%
where $J^{\mu }$ is the conserved current density, i.e.,
\begin{equation}
J^{\mu }=i\left( D^{\mu }\phi \right) ^{\ast }\phi -i\phi ^{\ast }\left(
D^{\mu }\phi \right) \text{.}
\end{equation}

Similarly, the field equation for the Higgs sector becomes
\begin{equation}
D_{\alpha }D^{\alpha }\phi +e^{2}\Psi ^{2}\phi +\frac{\partial U}{\partial
\phi ^{\ast }}=0\text{,}
\end{equation}%
whereas the ones for the fields $\Psi$ and $\chi$ are, respectively,
\begin{equation}
\partial _{\alpha }\left( G\partial ^{\alpha }\Psi \right) +2e^{2}\Psi
\left\vert \phi \right\vert ^{2}+\frac{\partial U}{\partial \Psi }=0\text{,}
\end{equation}
\begin{equation}
\square \chi +\left( \frac{1}{4}F_{\mu \nu }F^{\mu \nu }-\frac{1}{2}\partial
_{\mu }\Psi \partial ^{\mu }\Psi \right) \frac{\partial G}{\partial \chi }+%
\frac{\partial U}{\partial \chi }=0\text{,}
\end{equation}%
where $\square$  stands for the d'Alembertian operator.

Along  this manuscript, we are interested in stationary configurations. In this situation, the Gauss and Amp\`{e}re laws can be written, respectively, as
\begin{eqnarray}
&\displaystyle \partial_{j}\left( G\partial_{j}A_{0}\right)+\kappa B = 2e^{2}A_{0}\left\vert \phi \right\vert^{2}\text{,} & \label{Gagv1}\\[0.3cm]
&\displaystyle \partial_{k}\left( GB\right) +\kappa \partial _{k}A_{0}=-e\epsilon_{kj}J_{j}\text{,} & \label{Gagv2}
\end{eqnarray}%
where we have defined the magnetic field  {as} $B=F_{12}=\epsilon
_{ij}\partial _{i}A_{j}$. The presence of the Chern-Simons term in the Gauss
law allows us to conclude that the resulting solutions possess a nonvanishing electric charge and magnetic flux.

Similarly, the stationary equation for the Higgs field  {reads}%
\begin{equation}
D_{i}D_{i}\phi -e^{2}\Psi ^{2}\phi -\frac{\partial U}{\partial \phi ^{\ast }}%
=0\text{,}  \label{Gagv3}
\end{equation}%
while  the ones for $\Psi$ and $\chi$  assume the forms
\begin{eqnarray}
&\displaystyle \partial _{i}\left( G\partial _{i}\Psi \right) -2e^{2}\Psi \left\vert \phi \right\vert ^{2}-\frac{\partial U}{\partial \Psi }=0\text{,}&  \label{Gagv4} \\[0.2cm]
&\displaystyle \partial _{i}\partial _{i}\chi -\frac{1}{2}\left[ B^{2}-\left( \partial_{i}A_{0}\right) ^{2}+\left( \partial _{i}\Psi \right) ^{2}\right] \frac{\partial G}{\partial \chi }-\frac{\partial U}{\partial \chi }=0\text{,}&\quad\;\;
\label{Gagv5}
\end{eqnarray}%
respectively.

\subsection{The BPS formalism}

We focus our attention on those solutions satisfying a particular set of differential equations, i.e., the so-called BPS or self-dual ones. These equations arise through the implementation of the BPS prescription in the total energy of the model (\ref{1}), the starting-point being the expression for the stationary energy density, i.e.,
\begin{eqnarray}
\varepsilon &=&\frac{1}{2}GB^{2}+\left\vert D_{i}\phi \right\vert ^{2}+\frac{%
1}{2}G\left( \partial _{i}A_{0}\right) ^{2}+\frac{1}{2}G\left( \partial
_{i}\Psi \right) ^{2}  \notag \\[0.2cm]
&&+e^{2}\left( A_{0}\right) ^{2}\left\vert \phi \right\vert ^{2}+e^{2}\Psi
^{2}\left\vert \phi \right\vert ^{2}+\frac{1}{2}\left( \partial _{i}\chi
\right) ^{2}+U\text{,}  \label{eqen01}
\end{eqnarray}%
from which one gets the following boundary conditions:
\begin{eqnarray}
&&\displaystyle\lim_{\left\vert \mathbf{x}\right\vert \rightarrow \infty }%
\sqrt{G}B=0\text{ \ \ and \ \ }\lim_{\left\vert \mathbf{x}\right\vert
\rightarrow \infty }D_{i}\phi =0\text{,}  \label{eqcd01} \\[0.2cm]
&&\displaystyle\lim_{\left\vert \mathbf{x}\right\vert \rightarrow \infty }%
\sqrt{G}\partial _{i}A_{0}=0\text{ \ \ and \ \ }\lim_{\left\vert \mathbf{x}%
\right\vert \rightarrow \infty }\sqrt{G}\partial _{i}\Psi =0\text{,}
\label{eqcd02} \\[0.2cm]
&&\displaystyle\lim_{\left\vert \mathbf{x}\right\vert \rightarrow \infty
}A_{0}=0\text{ \ \ and \ \ }\lim_{\left\vert \mathbf{x}\right\vert
\rightarrow \infty }\Psi =0\text{,}  \label{eqcd03} \\[0.2cm]
&&\displaystyle\lim_{\left\vert \mathbf{x}\right\vert \rightarrow \infty
}\chi =0\text{ \ \ and \ \ }\lim_{\left\vert \mathbf{x}\right\vert
\rightarrow \infty }U=0\text{,}  \label{eqcd04}
\end{eqnarray}
which must be satisfied in order to ensure a finite energy.

The corresponding total energy reads as
\begin{equation}
E=\int d^{2}\mathbf{x}~\varepsilon \text{,}  \label{energi01}
\end{equation}%
which, after some algebraic manipulations, can be written in the form (here, we have introduced the auxiliary functions $\Sigma \equiv \Sigma \left(\phi, \Psi \right)$ and $F_{i}\equiv F_{i}\left( \chi \right)$, for the sake of convenience)
\begin{eqnarray}
E &=&\int d^{2}\mathbf{x}\left[ \frac{\left( GB\mp \Sigma \right) ^{2}}{2G}%
+\left\vert D_{\pm }\phi \right\vert ^{2}\mp \frac{1}{2}\epsilon
_{ij}\partial _{i}J_{j}\right.  \notag \\[0.2cm]
&&\hspace{1cm}+\frac{G}{2}\left( \partial _{i}A_{0}\mp \partial _{i}\Psi
\right) ^{2}+e^{2}\left( A_{0}\mp \Psi \right) ^{2}\left\vert \phi
\right\vert ^{2}  \notag \\[0.2cm]
&&\hspace{1cm}+\frac{1}{2}\left( \partial _{i}\chi \mp \epsilon _{ij}\frac{%
\partial F_{j}}{\partial \chi }\right) ^{2}\pm \epsilon _{ij}\partial
_{i}F_{j}  \notag \\[0.2cm]
&&\hspace{1cm}\pm B\left( \Sigma +e\left\vert \phi \right\vert ^{2}+\kappa
\Psi \right) \pm \partial _{j}\left( \Psi G\partial _{j}A_{0}\right)  \notag
\\[0.2cm]
&&\hspace{1cm}\left. +U-\frac{\Sigma ^{2}}{2G}-\frac{1}{2}\left( \frac{%
\partial F_{j}}{\partial \chi }\right) ^{2}\right] \text{,}  \label{energi02}
\end{eqnarray}%
where we have used not only the well-known identity
\begin{equation}
\left\vert D_{i}\phi \right\vert ^{2}=\left\vert D_{\pm }\phi \right\vert
^{2}\pm eB\left\vert \phi \right\vert ^{2}\mp \frac{1}{2}\epsilon
_{ij}\partial _{i}J_{j}\text{,}
\end{equation}%
but also the Gauss' law (\ref{Gagv1}).

Now, to continue with the implementation of the BPS formalism, it is useful to observe that, in view of the conditions (\ref{eqcd01}) and (\ref{eqcd02}), those integrations involving total derivatives do not contribute to the calculation of the total energy. In addition, the factor which multiplies the magnetic field allows us to fix the function $\Sigma$ by setting it as equal to $ev^{2}$, then we get
\begin{equation}
\Sigma =e\left( v^{2}-\left\vert \phi \right\vert ^{2}\right) -\kappa \Psi
\text{.}
\end{equation}

Furthermore, by setting the last row of the Eq. (\ref{energi02}) as equal to $0$, one gets the BPS potential
\begin{equation}
U=\frac{\left[ e\left( v^{2}-\left\vert \phi \right\vert ^{2}\right) -\kappa
\Psi \right] ^{2}}{2G}+\frac{1}{2}\left( \frac{\partial F_{j}}{\partial \chi
}\right) ^{2}\text{,}  \label{p}
\end{equation}%
where the first term in the right-hand side (with $G=1$) corresponds to the
potential which allows the standard Maxwell-Chern-Simons-Higgs model to be
self-dual \cite{mcshv}. In our enlarged case, the  {last} term stands
for the potential related to the field $\chi $, with $F_{i}(\chi)$ being
called a \textit{superpotential}.

In view of the considerations above, the total energy (\ref{energi02}) can be written as
\begin{equation}
E=E_{BPS}+E_{\text{quad}}\geq E_{BPS}\text{,}  \label{energi03}
\end{equation}
where $E_{BPS}$  is the Bogomol'ny bound given by
\begin{equation}
E_{BPS}=\pm ev^{2}\int d^{2}\mathbf{x~}B\pm \int d^{2}\mathbf{x~}\epsilon
_{ij}\partial _{i}F_{j}\geq 0\text{,}  \label{energi04}
\end{equation}%
and $E_{\text{quad}}$ represents the contribution of all the quadratic terms,  i.e.,
\begin{eqnarray}
E_{\text{quad}} &=&\int d^{2}\mathbf{x}\left\{ \frac{\left[ GB\mp e\left(
v^{2}-\left\vert \phi \right\vert ^{2}\right) \pm \kappa \Psi \right] ^{2}}{%
2G}\right.  \notag \\
&&\hspace{0.5cm}+\left\vert D_{\pm }\phi \right\vert ^{2}+\frac{1}{2}\left(
\partial _{i}\chi \mp \epsilon _{ij}\frac{\partial F_{j}}{\partial \chi }%
\right) ^{2}  \notag \\
&&\hspace{0.5cm}\left. +\frac{G}{2}\left( \partial _{i}A_{0}\mp \partial
_{i}\Psi \right) ^{2}+e^{2}\left( A_{0}\mp \Psi \right) ^{2}\left\vert \phi
\right\vert ^{2}\right\} \text{.}\quad \quad  \label{energi05}
\end{eqnarray}

We observe that,  {when} $E_{\text{quad}}=0$, the total energy attains its lower-bound,  i.e., $E=E_{BPS}$. In such a situation, the fields satisfy the following set of differential equations:
\begin{eqnarray}
&\displaystyle D_{\pm }\phi =0\text{,} & \label{qbps1}\\[0.2cm]
&\displaystyle GB=\pm e\left( v^{2}-\left\vert \phi \right\vert ^{2}\right) \mp \kappa \Psi\text{,}&  \label{qbps2} \\[0.2cm]
&\displaystyle \partial _{i}\chi =\pm \epsilon _{ij}\frac{\partial F_{j}}{\partial \chi}\text{,} & \label{qbps3}\\[0.3cm]
&\displaystyle\partial _{i}\Psi =\pm \partial _{i}A_{0}\text{ \ \ and \ \ }\Psi =\pm A_{0}\text{,} &  \label{qbps4}
\end{eqnarray}
which are the self-dual BPS equations whose solutions provide solitonic configurations with an energy which saturates a lower-bound, therefore guaranteeing the stability of the corresponding structures. It is worthwhile to comment that the set of Eqs. (\ref{qbps1})-(\ref{qbps4}) is equivalent to that of Euler-Lagrange equations (\ref{Gagv2})-(\ref{Gagv5}) associated with Lagrangian density (\ref{1}) whether we consider the BPS potential (\ref{p}).

The last two BPS equations in Eq. (\ref{qbps4}) are identically satisfied by $\Psi =\pm A_{0}$, from which the remaining equations
 {assume the form}%
\begin{eqnarray}
&\displaystyle D_{\pm }\phi =0\text{,} & \label{bpsx01} \\[0.2cm]
&\displaystyle GB=\pm e\left( v^{2}-\left\vert \phi \right\vert ^{2}\right) -\kappa A_{0}\text{,} & \label{bpsx02}\\[0.2cm]
&\displaystyle \partial _{i}\chi =\pm \epsilon _{ij}\frac{\partial F_{j}}{\partial \chi }\text{,} & \label{bpsx03}
\end{eqnarray}
which must be studied together with
\begin{equation}
\partial _{j}\left( G\partial _{j}A_{0}\right) +\kappa
B=2e^{2}A_{0}\left\vert \phi \right\vert ^{2}\text{,}  \label{bpsx04}
\end{equation}%
which is the Gauss law  of the model.

\subsection{The vortex ansatz}

We look for vortex configurations via standard map for the Higgs field,%
\begin{equation}
\phi \left( r,\theta \right) =vg(r)e^{in\theta }\text{,}  \label{anz01}
\end{equation}%
where $r$ and $\theta $ are the polar coordinates and~the integer $n=\pm 1,$
$\pm 2,$ $\pm 3,$ $...$ represents the winding number of the  {final}
solutions.

In addition, for the gauge field, we set%
\begin{equation}
A_{i}=\varepsilon _{ij}\frac{x_{j}}{er^{2}}\left[ a(r)-n\right] \text{,}
\label{anz02}
\end{equation}%
where $\varepsilon _{ij}$ is the Levi-Civita tensor and $x_{j}$ expresses the cartesian coordinates. It is interesting to note that the combination between the Eqs. (\ref{bpsx04}) and (\ref{anz02}) leads to the conclusion that the electric potential must depend on the radial coordinate only, i.e., $A_{0}=A_{0}(r)$.

Now, based on the Eqs. (\ref{anz01}) and (\ref{anz02}), it is reasonable to suppose that the scalar field $\chi$ and the superpotential $F_{i}$  can be parametrized as%
\begin{equation}
\chi =\chi (r)\text{ \ \ and \ \ }F_{i}=-\varepsilon _{ij}\frac{x_{j}}{r^{2}}%
W\left( \chi \right) \text{,}
\end{equation}%
 {while the field} profiles are {{expected to be}} regular
functions satisfying the following boundary conditions at $r=0$:%
\begin{equation}
g(0)=0\text{, \ \ }a(0)=n\text{ \ \ and \ \ }A_{0}(0)=cte\text{,}  \label{b1}
\end{equation}%
\begin{equation}
\chi (0)=\chi _{0}\text{ \ \ and \ \ }W(\chi _{0})=W_{0}\text{,}
\end{equation}%
and (here, prime denotes the derivative with respect to $r$)
\begin{equation}
g(\infty )=1\text{, \ \ }a(\infty )=0\text{ \ \ and \ \ }A_{0}^{\prime
}(\infty )=0\text{,}  \label{b2}
\end{equation}%
\begin{equation}
\chi(\infty )=\chi _{\infty }\text{ \ \ and \ \ }W(\chi _{\infty})= W_{\infty} \text{,}
\end{equation}%
which hold in the asymptotic limit $r\rightarrow \infty $ .

It proves useful to write the rotationally symmetric expressions for both the magnetic and electric fields, i.e.,
\begin{equation}
B(r)=-\frac{1}{er}\frac{da}{dr}\text{ \ \ and \ \ }E(r)=-\frac{dA_{0}}{dr}%
\text{,}  \label{mf}
\end{equation}%
which were obtained directly from the Eq. (\ref{anz02}).

By using the boundary conditions defined previously, we perform the integrals  which appear in the Eq. (\ref{energi04}) in order to calculate the value of the total BPS energy explicitly. As a result, we find that the Bogomol'nyi bound for the chiral Maxwell-Higgs electrodynamics reads
\begin{equation}
E_{BPS}=\pm \left( 2\pi v^{2}n+2\pi \Delta W\right) \geq 0\text{,}
\label{te1}
\end{equation}%
with $\Delta W=W_{\infty }-W_{0}$. In the expression above, the upper (lower) sign holds for positive (negative) values of both $n$ and $\Delta W$. In what follows, we analyze the case with $n>0$ and $\Delta W>0$ only, for the sake of simplicity.

The magnetic flux $\Phi$ can be calculated as
\begin{equation}
\Phi =\int B\,d^{2}\mathbf{x}=\frac{2\pi }{e}n\text{,}
\end{equation}%
while the total electric charge $Q$ of the chiral vortices is obtained as the integral of the Gauss law,  therefore reading
\begin{equation}
Q=\kappa \Phi \text{.}
\end{equation}

It is well known that the presence of the Chern-Simons term generates a
nonvanishing quantized angular momentum,  i.e.,
\begin{equation}
J=-\int \varepsilon _{ij}x_{i}T_{0j}\,d^{2}\mathbf{x}=\frac{\pi \kappa }{%
e^{2}}n^{2}=\frac{1}{4\pi }Q\Phi \text{,}
\end{equation}%
which has precisely the same value as the self-dual vortices of the pure Chern-Simons-Higgs model \cite{cshv}. This result does not depend on the function $G(\chi )$ whenever it satisfies the finite energy conditions given in  the Eqs. (\ref{eqcd01}) and (\ref{eqcd02}). The quantized angular momentum tells us that the chiral vortices also have an anyonic behavior due to the presence of the Chern-Simons term \cite{ref24}.

The rotationally symmetric version of the BPS Eqs. (\ref{bpsx01}) and (\ref{bpsx02}) and of the Gauss law (\ref{bpsx04}) can be written, respectively, as
\begin{equation}
\frac{dg}{dr}=\frac{ag}{r}\text{,}  \label{bps2}
\end{equation}%
\begin{equation}
GB=ev^{2}\left( 1-g^{2}\right) -\kappa A_{0}\text{,}  \label{bps1}
\end{equation}%
\begin{equation}
\frac{1}{r}\frac{d}{dr}\left( rG\frac{dA_{0}}{dr}\right) +\kappa
B=2e^{2}v^{2}g^{2}A_{0}\text{,}  \label{gl1}
\end{equation}%
while the Eq. (\ref{bpsx03}) can be verified to assume the form
\begin{equation}
\frac{d\chi }{dr}=\frac{1}{r}\frac{dW}{d\chi }\text{,}  \label{sf}
\end{equation}%
which, as we discuss separately below, plays a central role in the construction of the BPS chiral vortices.

Now, whether we use the BPS expressions, the Eq. (\ref{eqen01}) for the BPS energy density can be written as
\begin{equation}
\varepsilon _{BPS}=\varepsilon _{G}+\varepsilon _{\chi }\text{,}
\label{ed1xcp}
\end{equation}%
where $\varepsilon _{G}$ represents the contribution due to the
vortex alone,  i.e.,
\begin{equation}
\varepsilon _{G}=GB^{2}+G\left( \frac{dA_{0}}{dr}\right)
^{2}+2v^{2}e^{2}g^{2}\left( A_{0}\right) ^{2}+2v^{2}\frac{a^{2}g^{2}}{r^{2}}%
\text{,}  \label{edG}
\end{equation}%
while $\varepsilon _{\chi }$ corresponds to that exclusively related to the source field $\chi $,
\begin{equation}
\varepsilon _{\chi }=\left( \frac{d\chi }{dr}\right) ^{2}=\frac{1}{r^{2}}%
\left( \frac{dW}{d\chi }\right) ^{2}\text{.}  \label{edX}
\end{equation}

It is interesting to perceive that Eq. (\ref{sf}) for the field $\chi$ does not contain the other sectors of the model, therefore depending solely on the form of the superpotential $W(\chi)$. This fact allows us to solve it separately by choosing the superpotential ${W}(\chi)$ adequately. Once $\chi$ is known, we must selected the functional form of $G(\chi)$ and then proceed to the resolution of the set formed by the Eqs. (\ref{bps2}), (\ref{bps1}) and (\ref{gl1}). The next Section is devoted to this aim.

%\bigskip\medskip

\section{Some chiral vortices in the Maxwell-Higgs scenario\label{MCSHsc}}

We now investigate the construction of specific BPS scenarios and their respective numerical solutions. In this sense, we  first choose the superpotential as
\begin{equation}
W(\chi )=\chi -\frac{1}{3}\chi ^{3}\text{,}  \label{w}
\end{equation}%
which was used recently as an attempt to understand planar skyrmion-like
solitons \cite{2324} and the behavior of massless Dirac fermions in a
skyrmion-like background \cite{25}.

In view of this choice, the first-order equation (\ref{sf}) for the source
field can be written in the form%
\begin{equation}
\frac{d\chi }{dr}=\pm \frac{1}{r}\left( 1-\chi ^{2}\right) \text{,}
\label{w1}
\end{equation}%
whose exact solution reads%
\begin{equation}
\chi (r)=\pm \frac{r^{2}-r_{0}^{2}}{r^{2}+r_{0}^{2}}\text{,}  \label{ssf}
\end{equation}%
where $r_{0}$ represents an arbitrary positive constant such that $\chi(r_{0})= 0$. We note that this solution attains the boundary values $\chi_{0}=\mp 1$ and $\chi _{\infty }=\pm 1$. Our analysis only considers the upper sign, as previously commented after the Eq. (\ref{te1}).

In the sequence, we choose a particular expression for the dielectric function $G(\chi)$ which characterizes the chiral medium. In the following sections, we separate our study into two different scenarios.

%\bigskip \medskip

\subsection{Abrikosov-Nielsen-Olesen-like chiral vortices\label{mcshcc2}}

We consider the dielectric function $G(\chi)$ as
\begin{equation}
G(\chi )=\frac{1}{\chi ^{2}}=\frac{\left( r^{2}+r_{0}^{2}\right) ^{2}}{%
\left( r^{2}-r_{0}^{2}\right) ^{2}}\text{,}  \label{tg22}
\end{equation}%
where we have already used the Eq. (\ref{ssf}) for $\chi\left(r\right)$. We note that $r_{0}=0$ leads to $G\equiv 1$. As a consequence, the set of equations (\ref{bps2}), (\ref{bps1}) and (\ref{gl1}) correspond to those of the  standard MCSH electrodynamics. Furthermore, for $r_{0}>0$, the dielectric function behaves as $G(r=0)=1$ and $G(r\rightarrow \infty )\rightarrow 1$, which means that the medium does not change sensibly the behavior of the field profiles around the boundary conditions, i.e., they are expected to exhibit a behavior which is similar to that of the BPS solutions of the MCSH model. Nevertheless, the numerical analysis reveals that, along the radial coordinate, the dielectric function can alter the shape of the field profiles in a significant way.

We now use (\ref{tg22}) to rewrite the BPS equations (\ref{bps2}) and (\ref{bps1}) as
\begin{equation}
\frac{dg}{dr}=\frac{ag}{r}\text{,}  \label{b7}
\end{equation}%
\begin{equation}
B=-\frac{1}{er}\frac{da}{dr}=\frac{\left( r^{2}-r_{0}^{2}\right) ^{2}}{%
\left( r^{2}+r_{0}^{2}\right) ^{2}}\left[ ev^{2}\left( 1-g^{2}\right)
-\kappa A_{0}\right] \!\text{,}  \label{b8}
\end{equation}%
while the Gauss law (\ref{gl1}) assumes the form
\begin{equation}
\frac{1}{r}\frac{d}{dr}\left[ \frac{\left( r^{2}+r_{0}^{2}\right) ^{2}r}{%
\left( r^{2}-r_{0}^{2}\right) ^{2}}\frac{dA_{0}}{dr}\right] +\kappa
B=2e^{2}v^{2}g^{2}A_{0}\text{.}  \label{b9}
\end{equation}

Here, we pay attention to  the fact that, at $r=r_{0}$, the function $G(\chi)$ in (\ref{tg22}) diverges. Consequently, to attain a finite energy density (\ref{edG}), both the magnetic and electric fields must converge to zero  at $r=r_{0}$ faster than $G(\chi)$ diverges to counterbalance such a singularity. The numerical simulations show that it is exactly what happens, this way manifesting how the presence of the chiral medium can affect the field configurations.

\begin{figure}[t]
\centering\includegraphics[width=8.5cm]{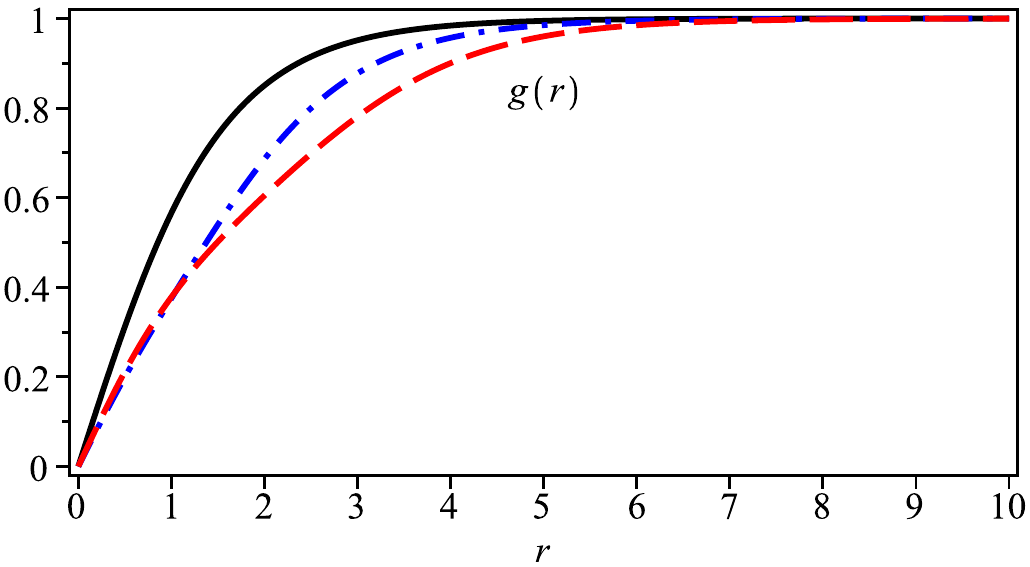}
\caption{Numerical solutions for the Higgs profile $g(r)$ obtained by solving the system of equations (\ref{b7}), (\ref{b8}) and (\ref{b9}) concerning the dielectric function (\ref{tg22}) for $r_{0}=1$ (dash-dotted blue line) and $r_{0}=2$ (dashed red line). The solid black line illustrates the solution for $G(\chi)\equiv 1$, which we have depicted here for the sake of comparison.}
\label{figg1}
\end{figure}

\begin{figure}[t]
\centering\includegraphics[width=8.5cm]{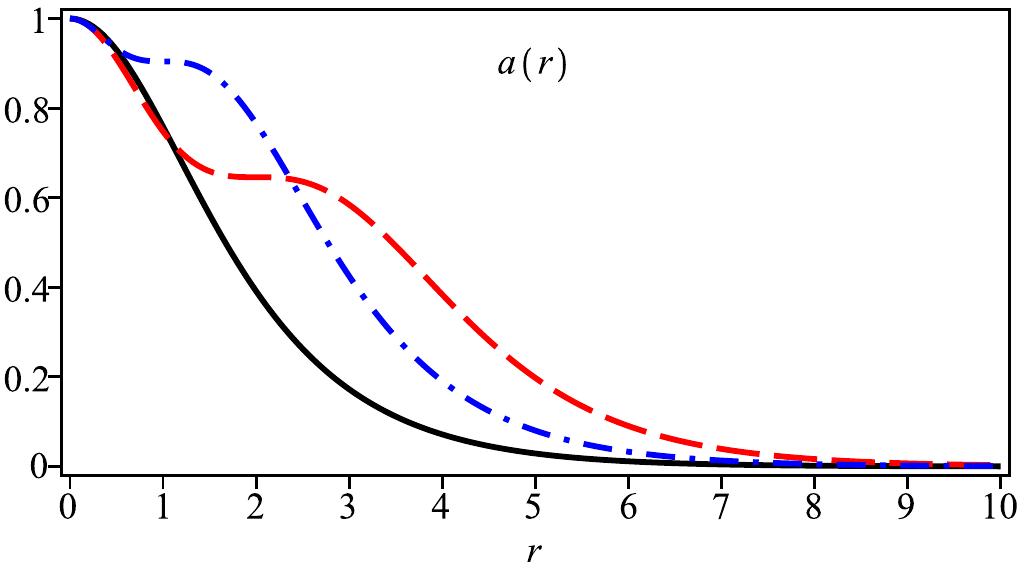}
\centering\includegraphics[width=8.5cm]{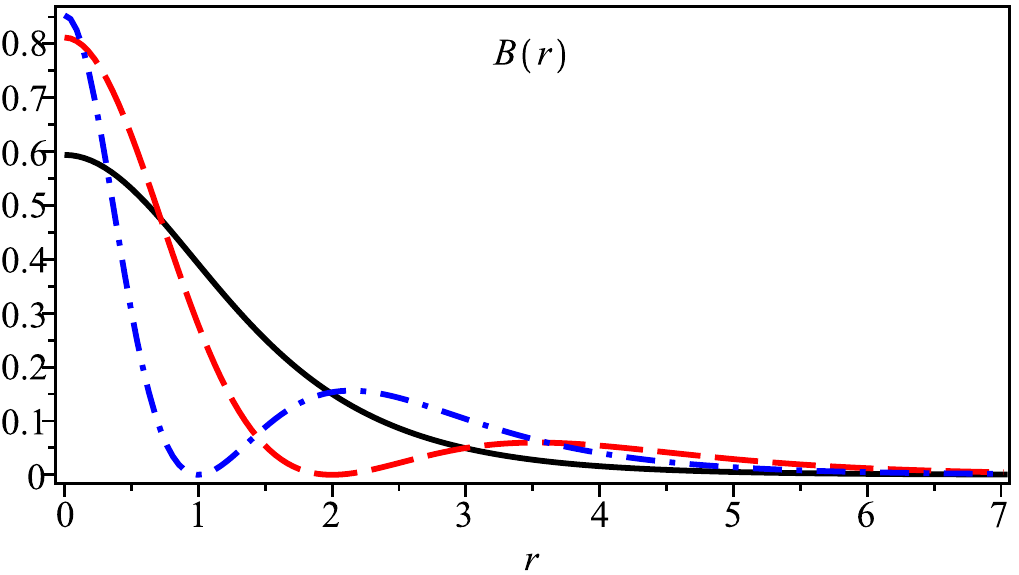}
\caption{Numerical solutions for the gauge profile function $a(r)$ (top) and the magnetic field $B(r)$ (bottom). Conventions as in the Fig. \ref{figg1}.} \label{figg2}
\end{figure}

We now verify how the dielectric medium changes the way the fields of the
model behave near the boundaries. By solving the Eqs. (\ref{b7}), (\ref{b8})
and (\ref{b9}) around the boundary values (\ref{b1}) and (\ref{b2}) such a
goal is attained. By considering $N>0$, we write below the behaviors near
the origin of the field profiles; this way, the function $g(r)$ reads
\begin{eqnarray}
g(r) &\approx &g_{N}r^{N}-\frac{eB_{0}g_{N}r^{N+2}}{4}+\frac{%
e^{2}v^{2}g_{N}^{3}r^{3N+2}}{4\left( N+1\right) ^{2}}  \notag \\[0.2cm]
&&+\left( \frac{eB_{0}}{8}-\frac{\kappa ^{2}}{16}+\frac{1}{r_{0}^{2}}\right)
\frac{eB_{0}g_{N}r^{N+4}}{4}\text{,}  \label{gaAc11}
\end{eqnarray}
whereas, the profile $a(r)$ behaves as
\begin{eqnarray}
a(r) &\approx &N-\frac{eB_{0}}{2}r^{2}-eB_{0}\left( \frac{\kappa ^{2}}{16}-%
\frac{1}{r_{0}^{2}}\right) r^{4}  \notag \\[0.2cm]
&&+\frac{e^{2}v^{2}g_{N}^{2}r^{2N+2}}{2\left( N+1\right) }\text{,}
\label{gaAc12}
\end{eqnarray}%
and respective magnetic field becomes
\begin{equation}
B(r)\approx B_{0}+4B_{0}\left( \frac{\kappa ^{2}}{16}-\frac{1}{r_{0}^{2}}%
\right) r^{2}-ev^{2}g_{N}^{2}r^{2N}\text{.}  \label{Bm_c1}
\end{equation}%
Besides, the scalar potential reads%
\begin{eqnarray}
A_{0}(r) &\approx &A_{0}(0)-\frac{\kappa B_{0}}{4}r^{2}+\frac{3\kappa B_{0}}{%
4}\left( -\frac{\kappa ^{2}}{48}+\frac{1}{r_{0}^{2}}\right) r^{4}  \notag \\%
[0.2cm]
&&+\frac{ev^{2}[2eA_{0}(0)+\kappa ]g_{N}^{2}r^{2N+2}}{4\left( N+1\right) ^{2}%
}\text{,}  \label{gaAc13}
\end{eqnarray}%
which leads to the electric field
\begin{eqnarray}
E(r) &\approx &\frac{\kappa B_{0}}{2}r+\left( \frac{\kappa ^{3}B_{0}}{16}-%
\frac{3\kappa B_{0}}{r_{0}^{2}}\right) r^{3}  \notag \\[0.2cm]
&&-\frac{ev^{2}[2eA_{0}(0)+\kappa ]g_{N}^{2}r^{2N+1}}{2\left( N+1\right) }%
\text{.}  \label{Ee_c1}
\end{eqnarray}

\begin{figure}[t]
\centering\includegraphics[width=8.5cm]{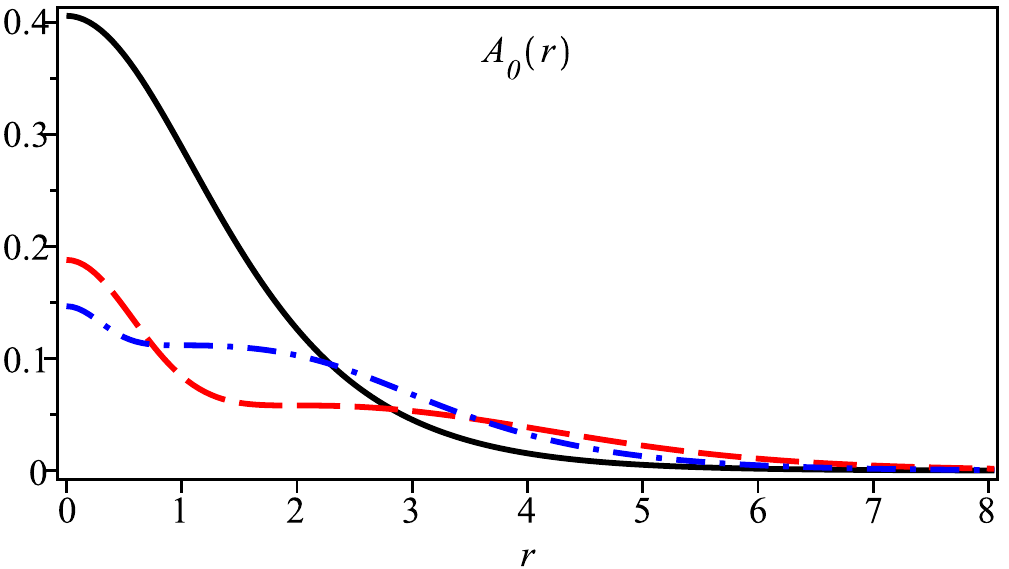} \centering%
\includegraphics[width=8.5cm]{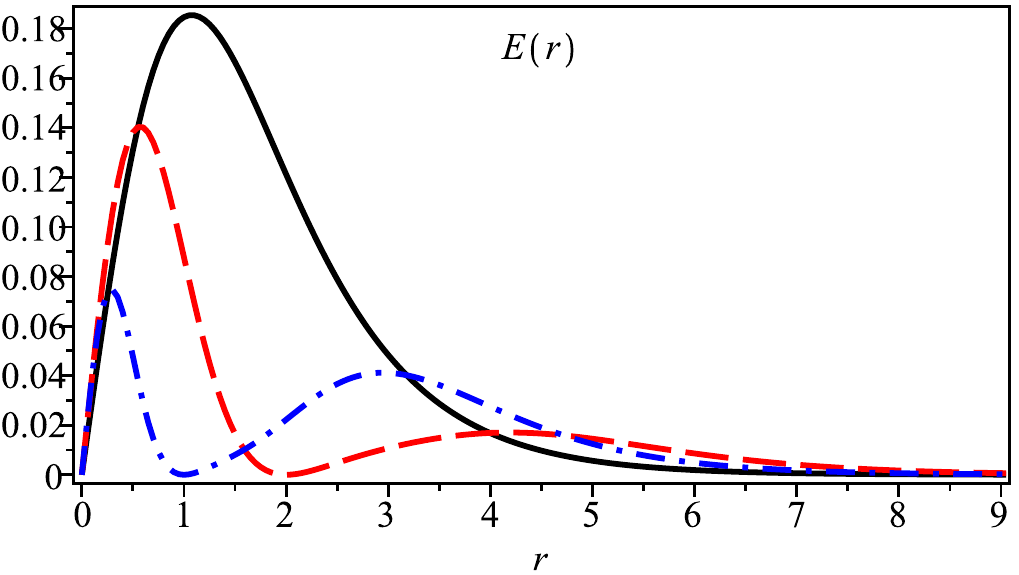}
\caption{Numerical solutions for the scalar potential $A_{0}(r)$ (top) and the electric field $E(r)$ (bottom). Conventions as in the Fig. \ref{figg1}.} \label{figg3}
\end{figure}

In the expressions above, the values of $g_{N}>0$, $B_{0}=B(r=0)$ and $A_{0}(0) =A_{0}(r=0)$ depend on the coupling constants and parameters of the model. In particular, the quantity $B_{0}$ is defined as
\begin{equation}
B_{0}=v^{2}e-\kappa A_{0}(0)  \label{B00}
\end{equation}%
and stands for the value of the magnetic field at the origin, while $A_{0}(0)$ is the amplitude of the scalar potential at $r=0$. For instance, for fixed $e=v=\kappa =1$ and $N=1$, the dependence on $r_{0}$ of both $B(0)$ and $A_{0}(0)$ are depicted in the Fig. \ref{figg5} below.

In addition, for the energy density $\varepsilon _{G}$, we obtain
\begin{eqnarray}
\varepsilon _{G}(r)
&=&B_{0}^{2}+2N^{2}v^{2}g_{N}^{2}r^{2N-2}+2e^{2}v^{2}(A_{0}(0))^{2}g_{N}^{2}r^{2N}
\notag \\[0.2cm]
&&+\left( \frac{3\kappa ^{2}B_{0}^{2}}{4}-\frac{4B_{0}^{2}}{r_{0}^{2}}%
\right) r^{2}  \notag \\[0.2cm]
&&-\left( N^{2}+2N+2\right) ev^{2}B_{0}g_{N}^{2}r^{2N}\text{,}  \label{EG_c1}
\end{eqnarray}%
from which the dependence of $\varepsilon _{G}(0)$ on $r_{0}$, again for $e=v= \kappa =1$ and $N=1$, is shown in the Fig. \ref{figg5}.

In general, when very close to the origin, the behaviors of the profiles resemble those of BPS solutions inherent to the MCSH electrodynamics, as expected.

On the other hand, in the asymptotic limit $r\rightarrow \infty $, the fields behave as
\begin{eqnarray}
g(r) &\approx &1-C_{\infty }\frac{e^{-Mr}}{\sqrt{r}}\text{,} \\[0.2cm]
a(r) &\approx &MC_{\infty }\sqrt{r}e^{-Mr}\text{,} \\[0.2cm]
A_{0}(r) &\approx &\frac{MC_{\infty }}{e}\frac{e^{-Mr}}{\sqrt{r}}\text{,}
\end{eqnarray}%
therefore preserving the typical behavior which characterizes the Abrikosov-Nielsen-Olesen's vortices. Further, the parameter $M$ represents the BPS mass of the bosonic fields and is given by
\begin{equation}
M=\frac{1}{2}\sqrt{8e^{2}v^{2}+\kappa ^{2}}-\frac{|\kappa |}{2}\text{,}
\end{equation}%
i.e., the exact definition inherent to the BPS vortices obtained in the Maxwell-Chern-Simons-Higgs model. Then, as previously pointed out, the dielectric function (\ref{tg22}) does not change sensibly how the fields approach the vacuum.

\begin{figure}[t]
\centering\includegraphics[width=8.5cm]{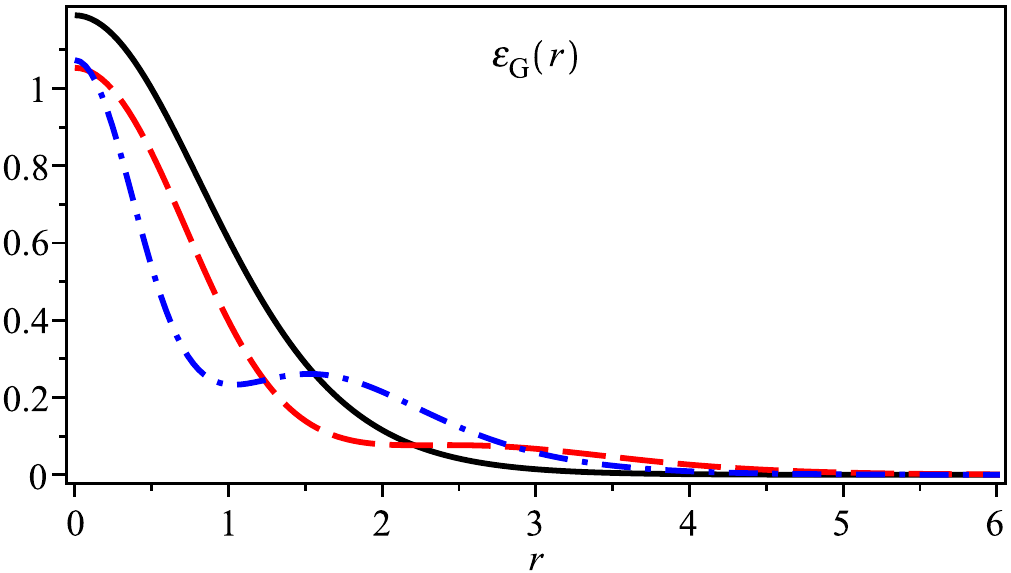} \centering
\caption{Numerical solutions for the energy density $\varepsilon_{G}(r)$ of the BPS configurations. Conventions as in the Fig. \ref{figg1}.} \label{figg4}
\end{figure}

%\bigskip\medskip

\subsubsection*{Numerical analysis -- ANO-like chiral vortices:}

\begin{figure}[t]
\centering\includegraphics[width=8.5cm]{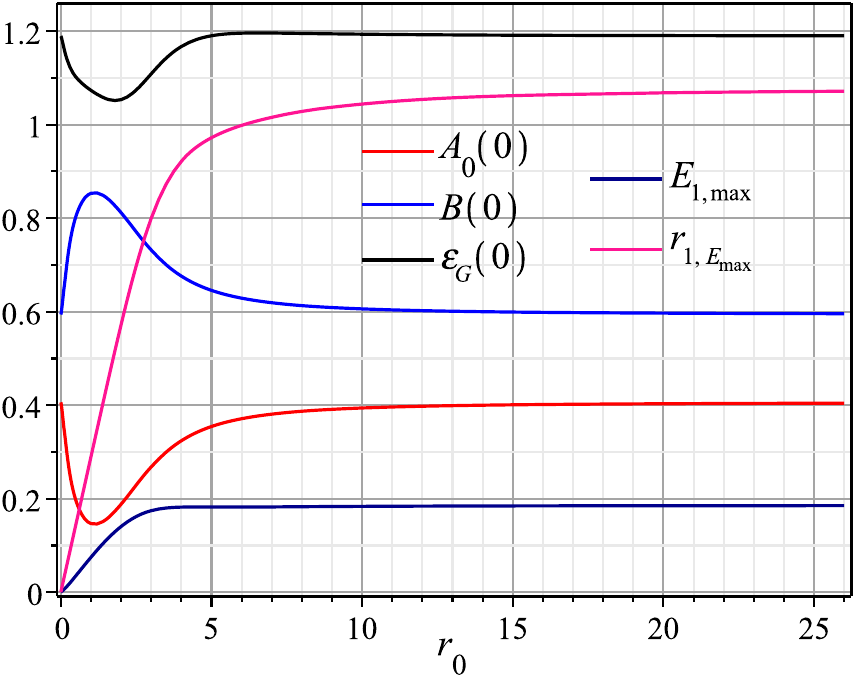}
\caption{Numerical results for the quantities $A_{0}(0)$, $B(0)$ and $\varepsilon_{G}(0)$ as functions of the dielectric parameter $r_{0}$. In this Figure, $E_{1,\text{max}}$ and $r_{1, E_{\text{max}}}$ represent the amplitude of the first maximum of the electric field and its position along the radial coordinate, respectively, also as functions of $r_{0}$. We have obtained these results for the dielectric function $G(\chi)$ as given in Eq. (\ref{tg22}). } \label{figg5}
\end{figure}

\begin{figure}[t]
\centering\includegraphics[width=8.4cm]{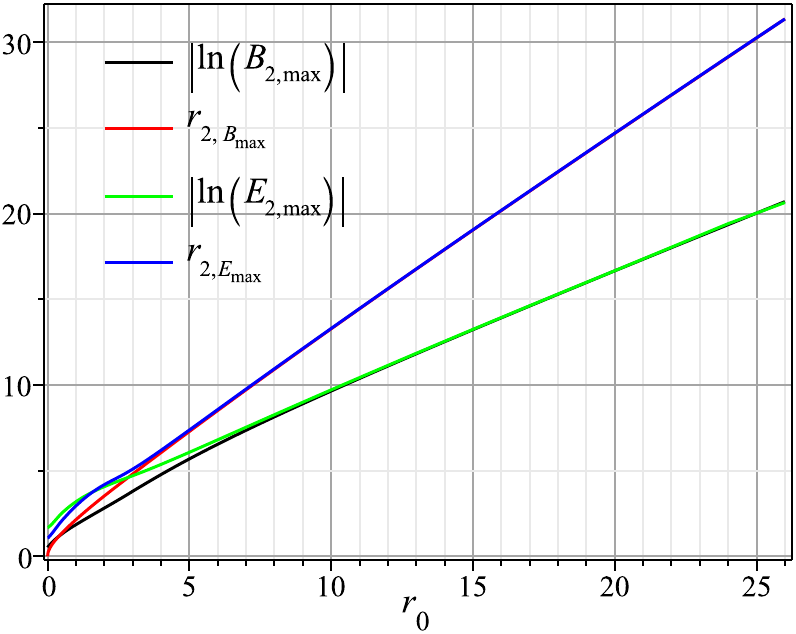}
\caption{Numerical results to the quantities $B_{2,\text{max}}$ and $E_{2,\text{max}}$ (i.e., the amplitudes of the second maximums of the magnetic and the electric fields, respectively), together with their respective positions along the radial coordinate (namely, $r_{2, B_{\text{max}}}$ and $r_{2, E_{\text{max}}}$, respectively). We have obtained these results for the dielectric function $G(\chi)$  as given in Eq. (\ref{tg22}).}
\label{figg6}
\end{figure}

Hereafter, we fix the constants of the model as $e=v=1$, $\kappa =1$, $n=1$,
and  {select} $r_{0}=1$ and $r_{0}=2$ to solve, through the finite-difference technique and according to the boundary conditions (\ref{b1}) and (\ref{b2}), the set defined by the Eqs. (\ref{b7}), (\ref{b8}) and (\ref{b9}),  from which we compare the new solutions with the ones attained when $G\equiv 1$. The figures \ref{figg1}, \ref{figg2}, \ref{figg3} and \ref{figg4} below show the numerical solutions for the profile functions $g(r)$,  $a(r)$ and the magnetic field $B(r)$, the scalar potential $A_{0}(r)$ and the electric $E(r)$, and  the energy distribution $\varepsilon _{G}$. The solid black line represents the solution for $G\equiv1$.  Further, in the figures \ref{figg5} and \ref{figg6}, we have depicted the dependence of $A_{0}(r=0)$, $B(r=0)$ and $\varepsilon _{G}(r=0)$ on the dielectric parameter $r_{0}$. We have also studied how the maximums (and their respective localizations along the radial axis) of both the electric and magnetic fields depend on the dielectric parameter.

The Higgs profile function $g(r)$ grows monotonically according to the boundary values (\ref{b1}) and (\ref{b2}), see the Fig. \ref{figg1}. The numerical analysis reveals that the size of the core inherent to these profiles becomes larger whether the dielectric parameter increases within the interval $0 \leq r_{0}< r^{\ast(1)}$. In particular, when $r_{0}=r^{\ast(1)}$, the core attains its maximum width.  On the other hand, for $r_{0}>r^{\ast (1)}$, the width of the core diminishes until it achieves the value obtained when $G\equiv 1$. Here, for instance, whether we define the width of the profile as $\approx 63.2\%$ of its height, then the maximum width of the profile is attained when $r^{\ast (1)}\approx 1.95$.

%%%%%%%%%%%%%%%%%%% Higgs field %%%%%%%%%%%%%%%%%%%%

%%%%%%%%%%%%%%%%%%% vector and magnetic field %%%%%%%%%%%%%%%%%%%%

Figure \ref{figg2} displays the solutions for both the gauge profile $a(r)$ and the magnetic field $B(r)$. The gauge profiles present a similar dependence with $r_{0}$ compared to the found for the Higgs sector. The novelty here is the development of a plateau around the point $r=r_{0}$ that becomes larger whenever the dielectric parameter increases. At the same time, the value of $a(r_{0})$ diminishes continuously. On the other hand, the magnetic field profiles do not vanish at the origin, whose amplitudes at $r=0$ vary with $r_{0}$ as shown by the blue line in Fig. \ref{figg5}.  Note that, for sufficiently large values of $r_{0}$, the value of $B(0)$  converges to that obtained when $G\equiv 1$. Furthermore, the magnetic field vanishing at $r=r_{0}$ develops a second maximum whose amplitude $B_{2,\text{max}}$ and its location $r_{2, B_{m_{ax}}}>r_0$, as functions of $r_0$, are both illustrated, respectively, by the black and red lines in the Fig. \ref{figg6}. Moreover, for $r_{0}$ sufficiently large, we observe that the second maximum decreases exponentially with $r_{0}$ (i.e., $\sim \exp (-0.687r_{0})$), while, in contrast, the respective location grows linearly with $r_{0}$.

%%%%%%%%%%%%%%%%%%% scalar potential and electric field %%%%%%%%%%%%%%%%%%%%

The scalar potential $A_{0}(r)$ at the origin is nonzero whose amplitude denoted by $A_{0}(0)$ depends on $r_{0}$ such as depicted by the red line in the Fig. \ref{figg5}. Along the radial coordinate, $A_{0}(r)$ behaves as the gauge profile $a(r)$ in the sense that it also develops a plateau centered at $r=r_{0}$; see the Fig. \ref{figg3}. Similarly, as it happens with the magnetic field, the resulting electric field also vanishes at $r=r_{0}$; consequently, it exhibits two different maximums: the first one $E_{1,\text{max}}$ located at $r_{1, E_{\text{max}}}$, which is within the interval $0\leq r<r_{0}$, and the second maximum $E_{2,\text{max}}$ positioned at $r_{2, E_{\text{max}}}$ on the region $r>r_{0}$. In contrast, the electric field without the dielectric function possesses a unique maximum located at $r_{1}^{\ast}=1.076$, whose amplitude is $\thickapprox 0.186$, see the black line in the bottom picture of the Fig. \ref{figg3}.  In Figure \ref{figg5}, both $E_{1,\text{max}}$ and $r_{1, E_{\text{max}}}$ are depicted as functions of $r_{0}$, whence we observe that for $r_{0}$ sufficiently large both converge to the values attained when $G(\chi)\equiv 1$. On the other hand, the green and the blue lines appearing in Fig. \ref{figg6} illustrate, respectively, how the amplitude $E_{2,\text{max}}$ and its localization $r_{2, E_{\text{max}}}$ vary with the dielectric parameter $r_{0}$. We notice that, for sufficiently large values of $r_{0}$, the second maximum decreases exponentially with $r_{0}$ (i.e., $\sim \exp (-0.687r_{0}$)), while its position increases linearly.

%%%%%%%%%%%%%%%%%%% energy density $\varepsilon_{G}$ %%%%%%%%%%%%%%%%%%%%

The numerical results for the energy density $\varepsilon_{G}(r)$ are shown in Fig \ref{figg4}. Such as it happens when $G(\chi)\equiv 1$, the value at $r=r_0$ denoted as $\varepsilon_{G}(0)$ is nonnull, but now it depends on the dielectric parameter $r_{0}$ such as depicted as the black line in the Fig. \ref{figg5}. Besides, we also observe that there is a value $r^{\ast}_{\text{en}}$ such that for $0<r_{0}< r_{\text{en}} ^{\ast}$ a local minimum arises around $r=r_{0}$. Simultaneously, a second maximum appears on the right-hand side of $r_{0}$. However, both extremes disappear for $r_{0}>r_{\text{en}}^{\ast }$, such that, for sufficiently large values of $r_{0}$, the profiles attain the same format as in the case with $G(\chi)\equiv 1$.

\begin{figure}[t]
\includegraphics[width=8.5cm]{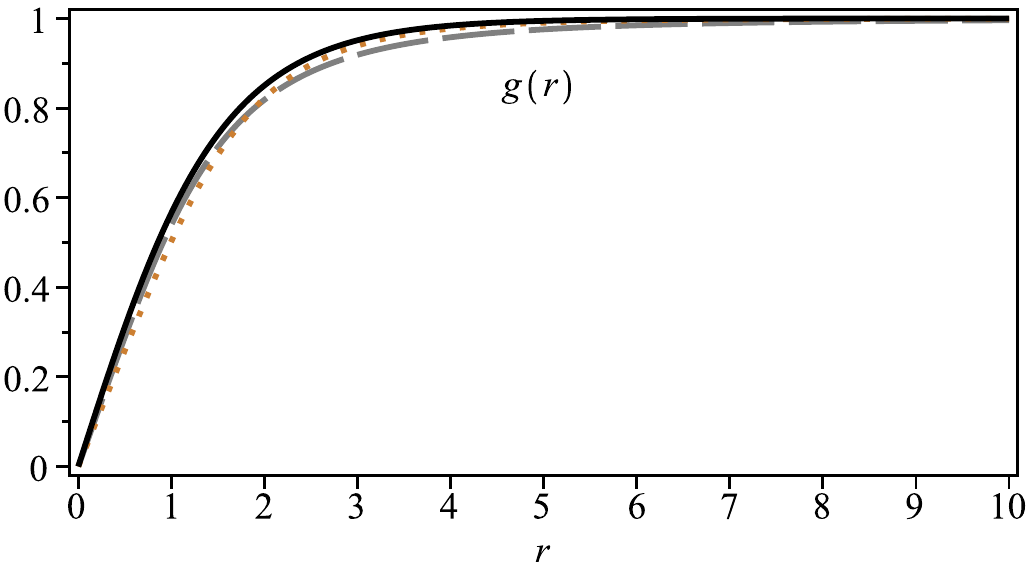}
\caption{Numerical solutions for the Higgs profile $g(r)$ obtained by solving the system of equations (\ref{b4}), (\ref{b5}) and (\ref{b6}) regarding the dielectric function (\ref{tg11})for $r_{0}=1$ (long-dashed gray line) and  $r_{0}=2$ (dotted orange line).  Again, the solid black line represents the solution for $G(\chi)\equiv 1$, which we have depicted here for the sake of comparison.}  \label{figg7}
\end{figure}

\subsection{Delocalized chiral vortices \label{fcmcsh1}}

In what follows, we choose the dielectric function as%
\begin{equation}
G(\chi )=\frac{1}{1-\chi ^{2}}=\frac{\left( r^{2}+r_{0}^{2}\right) ^{2}}{%
4r^{2}r_{0}^{2}}\text{,}  \label{tg11}
\end{equation}%
where we have already considered the solution (\ref{ssf}), which is finite along the radial axis but diverges at  $r=0$ and when $r\rightarrow \infty$. Therefore, to avoid the energy density (\ref{edG}) being singular, both the magnetic field and electric field must vanish at the boundaries more rapidly than the dielectric function diverges in such a way that the total energy converges to the finite value given by (\ref{te1}).

By using (\ref{tg11}), the self-dual equations (\ref{bps2}) and (\ref{bps1})
assume the form%
\begin{equation}
\frac{dg}{dr}=\frac{ag}{r}\text{,}  \label{b5}
\end{equation}%
\begin{equation}
B=-\frac{1}{er}\frac{da}{dr}=\frac{4r^{2}r_{0}^{2}}{\left(
r^{2}+r_{0}^{2}\right) ^{2}}\left[ ev^{2}\left( 1-g^{2}\right) -\kappa A_{0}%
\right] \!\text{,}  \label{b4}
\end{equation}%
while the Gauss law (\ref{gl1}) can be rewritten as
\begin{equation}
\frac{1}{r}\frac{d}{dr}\left[ \frac{\left( r^{2}+r_{0}^{2}\right) ^{2}}{%
4r_{0}^{2}r}\frac{dA_{0}}{dr}\right] +\kappa B=2e^{2}v^{2}g^{2}A_{0}\text{.}
\label{b6}
\end{equation}

We now investigate how the dielectric medium affects the behavior of the fields close to the boundary values given by the Eqs. (\ref{b1}) and (\ref{b2}). With this purpose in mind, we solve the equations (\ref{b5}), (\ref{b4}) and (\ref{b6}) around the boundaries. In this case, for $n=N>0$, we get that, near the origin, the Higgs profile behaves as
\begin{eqnarray}
g(r) &\approx &g_{N}r^{N}-\frac{eB_{0}g_{N}r^{N+4}}{4r_{0}^{2}}+\frac{%
2eB_{0}g_{N}r^{N+6}}{9r_{0}^{4}}  \notag \\[0.2cm]
&&+\frac{e^{2}v^{2}g_{N}^{3}r^{3N+4}}{\left( N+2\right) ^{2}r_{0}^{2}}\text{,%
}  \label{gaAc21}
\end{eqnarray}%
where $g_{N}>0$ and $B_{0}$ is defined as in  the Eq. (\ref{B00}). We observe that  the next-to-leading-order contribution is proportional to $r^{N+4}$, whereas the one in (\ref{gaAc11}) is to $r^{N+2}$.

\begin{figure}[t]
\includegraphics[width=8.5cm]{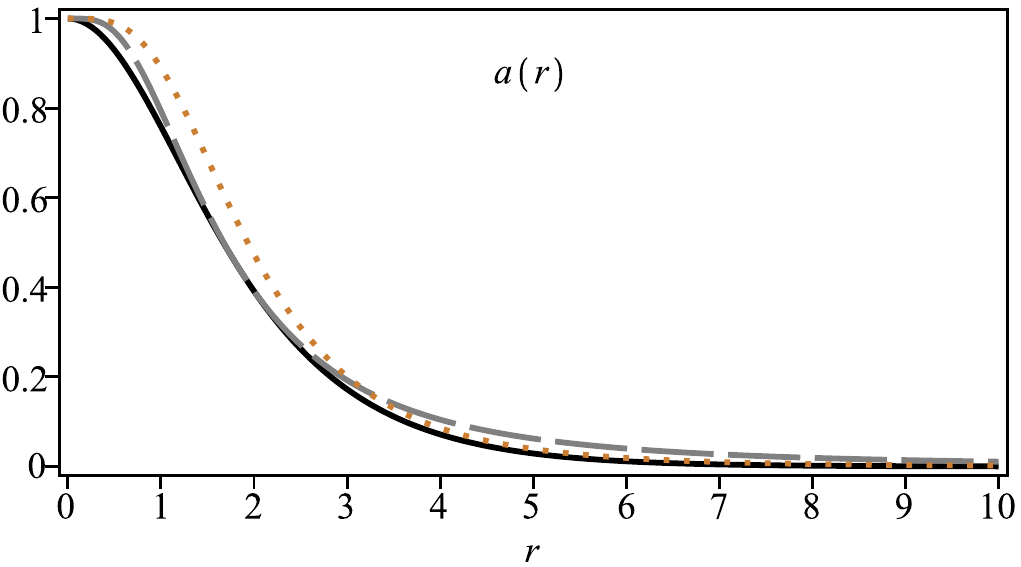} %
\includegraphics[width=8.5cm]{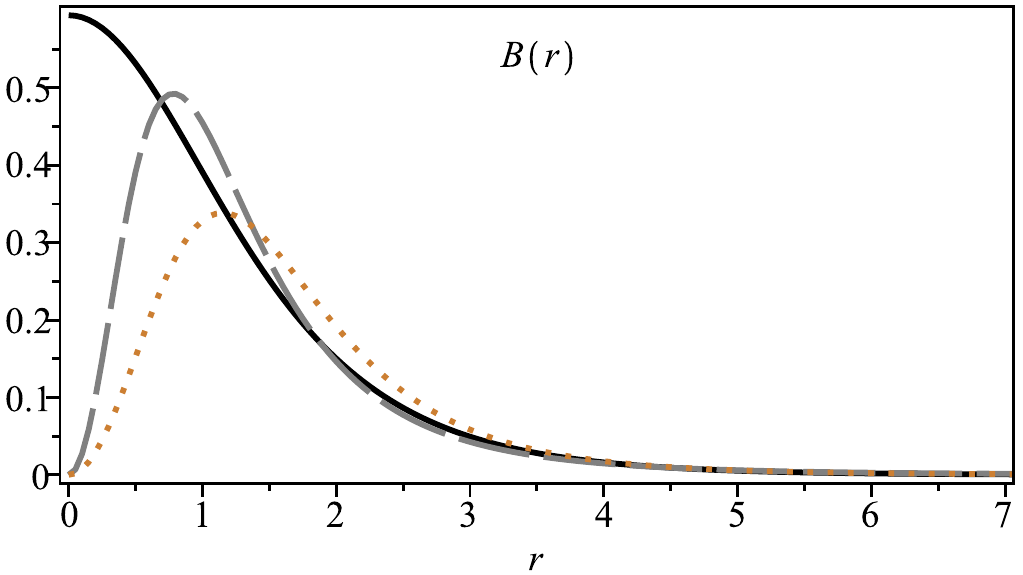}
\caption{Numerical solution for the gauge profile $a(r)$ (top) and the magnetic field $B(r)$ (bottom). Conventions as in the Fig. \ref{figg7}.} \label{figg8}
\end{figure}

In the meanwhile, the gauge profile $a(r)$ reads
\begin{equation}
a(r)\approx N-\frac{eB_{0}r^{4}}{r_{0}^{2}}+\frac{3eB_{0}r^{6}}{4r_{0}^{4}}+%
\frac{2e^{2}v^{2}g_{N}^{2}r^{2N+4}}{\left( N+2\right) r_{0}^{2}}\text{,}
\label{gaAc22}
\end{equation}%
from which we see that it converges faster to the value $N$ than the one in (\ref{gaAc12}). The corresponding magnetic field behaving as
\begin{eqnarray}
B(r) &\approx &\frac{4B_{0}}{r_{0}^{2}}r^{2}-\frac{8B_{0}}{r_{0}^{4}}r^{4}+%
\frac{12B_{0}}{r_{0}^{6}}r^{6}  \notag \\[0.2cm]
&&-\frac{4ev^{2}}{r_{0}^{2}}g_{N}^{2}r^{2N+2}+\frac{8ev^{2}}{r_{0}^{4}}%
g_{N}^{2}r^{2N+4}\text{,}  \label{Bm_c2}
\end{eqnarray}%
vanishes at $r=0$,  contrasting the one in (\ref{Bm_c1}) and acquiring a maximum outside from the origin; see Fig \ref {figg8}.

\begin{figure}[t]
\includegraphics[width=8.5cm]{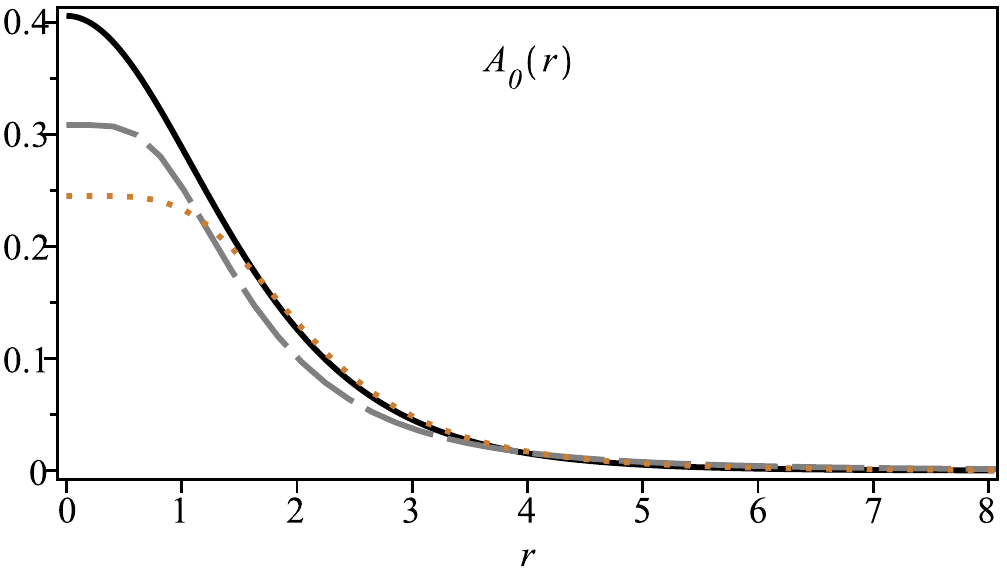} %
\includegraphics[width=8.5cm]{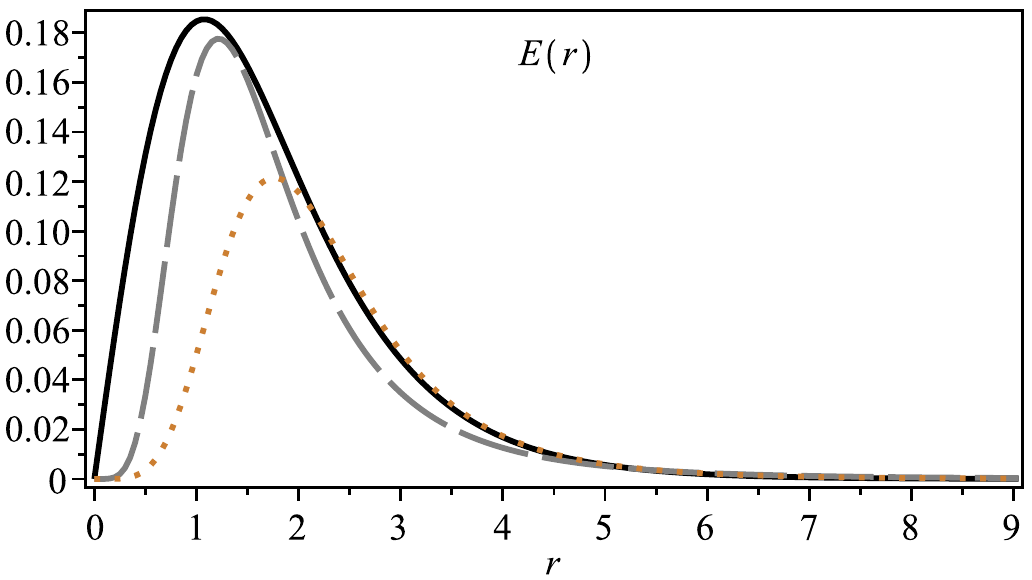}
\caption{Numerical solutions for the scalar potential $A_{0}(r)$ (top) and the electric field $E(r)$ (bottom). Conventions as in the Fig. \ref{figg7}.} \label{figg9}
\end{figure}

\begin{figure}[t]
\centering\includegraphics[width=8.6cm]{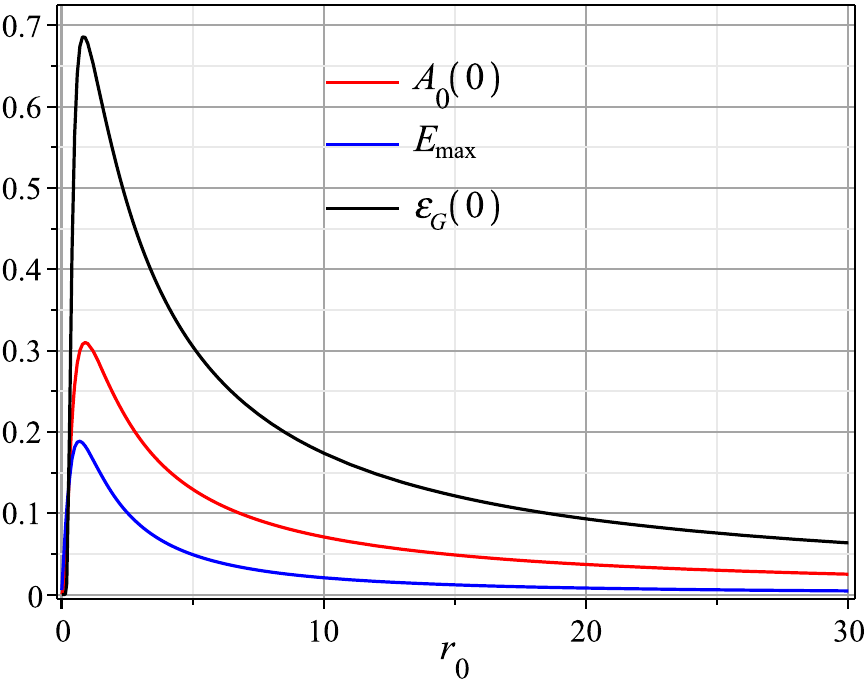}
\caption{Numerical results for the quantities $A_{0}(0)$, $\varepsilon_{G}(0)$, and $E_{\text{max}}$ (the last one represents the maximum value of the electric field). We have obtained these results for the dielectric function $G(\chi)$  as given in  the Eq. (\ref{tg11}).} \label{figg11}
\end{figure}

The scalar potential near the origin reads
\begin{eqnarray}
A_{0}(r) &\approx &A_{0}(0)-\frac{2\kappa B_{0}r^{6}}{3r_{0}^{4}}+\frac{%
5\kappa B_{0}r^{8}}{3r_{0}^{6}}  \notag \\[0.2cm]
&&+\frac{2e^{2}v^{2}A_{0}(0)g_{N}^{2}r^{2N+4}}{\left( N+1\right) \left(
N+2\right) r_{0}^{2}}+\frac{4ev^{2}{\kappa }g_{N}^{2}r^{2N+6}}{\left(
N+2\right) \left( N+3\right) r_{0}^{4}}  \notag \\[0.3cm]
&&-\frac{4e^{2}v^{2}A_{0}(0)g_{N}^{2}r^{2N+6}}{\left( N+1\right) \left(
N+3\right) r_{0}^{4}}\text{,}  \label{gaAc23}
\end{eqnarray}%
(where $A_0(0)$ represents the scalar potential value at $r=0$) whose next-to-leading-order contribution is proportional to $r^{6}$, whereas the one in (\ref{gaAc13}) is to $r^{2}$. Furthermore, for fixed $e=v=\kappa=1$ and $N=1$, the amplitude $A_{0}(r=0)$ vs. $r_{0}$ behaves differently from the one shown in  the Fig. \ref{figg5},  see the Fig. \ref{figg11}. Furthermore, the electric field is given by
\begin{eqnarray}
E(r) &\approx &\frac{4\kappa B_{0}r^{5}}{r_{0}^{4}}-\frac{40\kappa B_{0}r^{7}%
}{3r_{0}^{6}}-\frac{4e^{2}v^{2}A_{0}(0)g_{N}^{2}r^{2N+3}}{\left( N+1\right)
r_{0}^{2}}  \notag \\[0.2cm]
&&+\frac{8e^{2}v^{2}A_{0}(0)g_{N}^{2}r^{2N+5}}{\left( N+1\right) r_{0}^{4}}-%
\frac{8ev^{2}{\kappa }g_{N}^{2}r^{2N+5}}{\left( N+2\right) r_{0}^{4}}\text{,}
\label{Ee_c2}
\end{eqnarray}%
which converges faster to zero at the origin than that in Eq. (\ref{Ee_c1}).

Finally, near the origin, the energy density $\varepsilon _{G}(r)$ can be expressed as
\begin{eqnarray}
\varepsilon _{G}(r) &\approx
&2N^{2}v^{2}g_{N}^{2}r^{2N-2}+2e^{2}v^{2}A_{0}(0) g_{N}^{2}r^{2N}  \notag \\%
[0.3cm]
&&+\frac{4B_{0}^{2}}{r_{0}^{2}}r^{2}-\frac{8B_{0}^{2}}{r_{0}^{4}}r^{4}
\notag \\[0.2cm]
&&-\left( N^{2}+4N+8\right) \frac{ev^{2}B_{0}}{r_{0}^{2}}g_{N}^{2}r^{2N+2}%
\text{,}  \label{EG_c2}
\end{eqnarray}%
which contains more contributions associated with the dielectric parameter $r_{0}$ than the solution in the Eq. (\ref{EG_c1}). The amplitude of $\varepsilon_{G}(r)$ at origin, $\varepsilon_{G}(0)$, in terms of $r_{0}$ is depicted in the Fig. \ref{figg11}.

Thus, we observe that close to the origin, the behavior of the field profiles shows the strong influence of the dielectric function (\ref{tg11}) through the parameter $r_{0}$, resulting in very different behavior when compared with that of the BPS solutions of electrodynamics MCSH, as the numerical analysis below demonstrates.

\begin{figure}[t]
\centering\includegraphics[width=8.5cm]{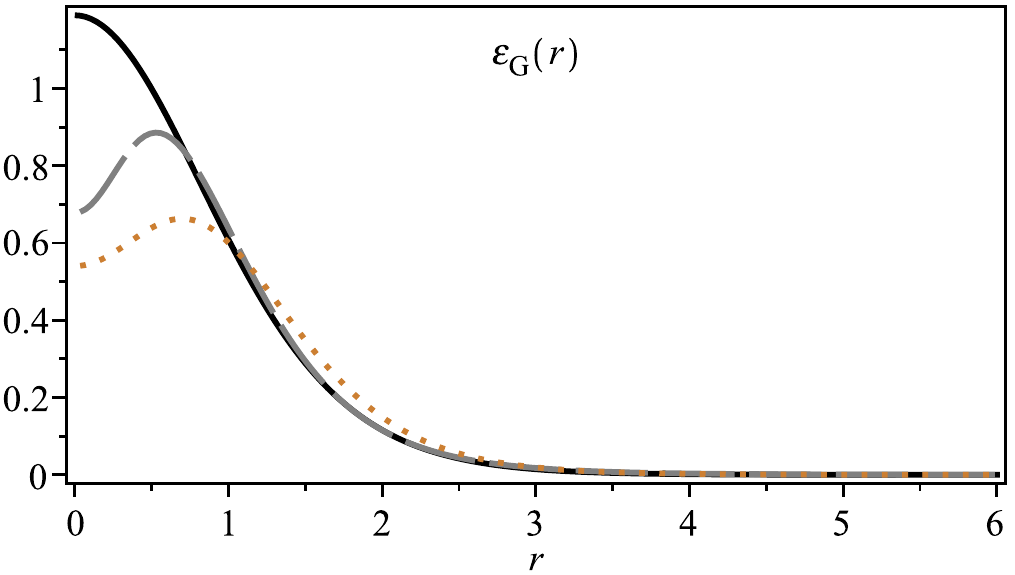} \centering
\caption{Numerical solutions for the energy density $\varepsilon_{G}(r)$ of the BPS configurations. Conventions as in the Fig. \ref{figg7}.} \label{figg10}
\end{figure}

In the limit $r\rightarrow \infty $, the profile functions read
\begin{eqnarray}
g(r) &\approx &1-C_{\infty }r^{-\Lambda }\text{,} \\[0.3cm]
a(r) &\approx &\Lambda C_{\infty }r^{-\Lambda }\text{,} \\[0.3cm]
A_{0}(r) &\approx &-\frac{\Lambda C_{\infty }}{2e\kappa \left( r_{0}\right)
^{2}}r^{-\Lambda }\text{,}
\end{eqnarray}%
where $C_{\infty }$ is a positive integration constant.  {Here,} we
have introduced the parameter $\Lambda $,  i.e.,%
\begin{equation}
\Lambda =1+\sqrt{1+8\left( evr_{0}\right) ^{2}}\text{,}  \label{ww1}
\end{equation}%
which reveals how the power-law decay depends on $r_{0}$. We here highlight that for $r\rightarrow \infty$, the presence of the dielectric medium ruled by $G(\chi)$ as defined in (\ref{tg11}) has radically changed the behavior of the field profiles, whose tails now decay by following a power-law, in contrast to the canonical behavior (i.e., the exponential decay one).

The literature shows that the London limit \cite{Babaev56} provides the behavior of the fields of a vortex via the full Ginzburg-Landau model, therefore predicting that the magnetic field varies monotonically and that, at large distances, is exponentially localized. However, very recently, it has been discovered that some vortex solutions have a delocalized magnetic field whose profiles decay slowly. These delocalized vortices whose tails follow a power-law decay have arisen during the study of the magnetic field delocalization in two-component superconductors \cite{Babaev57}, and also in the context of $k$-generalized Maxwell-Higgs electrodynamics \cite{jehp_andre}. Furthermore, such behavior has also been reported in diamagnetic vortices generated within a Chern-Simons theory \cite{Shaposhnikov58}.

\subsubsection*{Numerical analysis -- Delocalized chiral vortices:}

\begin{figure}[t]
\centering\includegraphics[width=8.5cm]{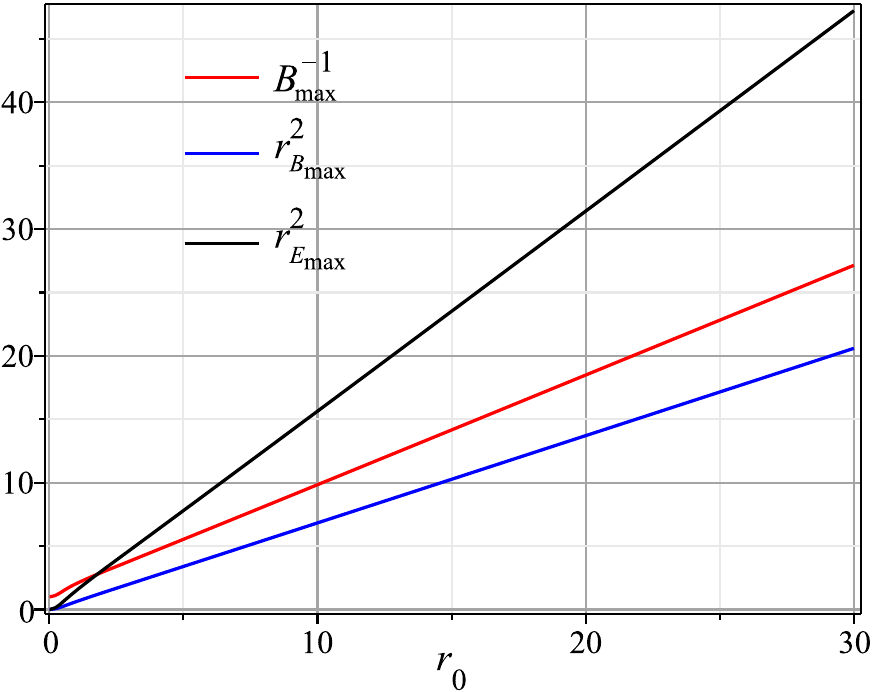}
\caption{Numerical results for the quantity $B_{\text{max}}$ (the maximum amplitude of the magnetic field) and $r_{B_{\text{max}}}$ (the position of $B_{\text{max}}$). Here, $r_{E_{\text{max}}}$ represents the point at which the
electric field achieves its maximum. We have obtained these results for the dielectric function $G(\chi )$ as given in the Eq. (\ref{tg11}).} \label{figg12}
\end{figure}

We again fix the constants of the model as $e=v=1$, $\kappa =1$, $n=1$, via which we solve the set defined by the Eqs.  (\ref{b5}), (\ref{b4}) and (\ref{b6}) according to the boundary conditions (\ref{b1}) and (\ref{b2}) through the finite-difference technique, for $r_{0}=1$ and $r_{0}=2$. We then compare the new solutions with the ones attained when $G\equiv1$ (which appear as the solid black lines in the plots below). The figures \ref{figg7}, \ref{figg8}, \ref{figg9} and \ref{figg10} depict, respectively, the numerical results for the Higgs profile $g(r)$, the vector gauge profile $a(r)$ and the magnetic field $B(r)$, the scalar potential $A_{0}(r)$ and the electric $E(r)$ field, and the energy distribution $\varepsilon _{G}$. Furthermore, in the figures \ref{figg11} and \ref{figg12}, we analyze the dependence of the amplitudes at $r=0$ of the scalar potential and energy density on the dielectric parameter $r_{0}$. We also look at how the maxima of the electric and magnetic fields (and their respective localizations along the radial coordinate) depend on the dielectric parameter.

%%%%%%%%%%%%%%%%%%%%%%%

%%%%%%%%%%%%%%%%%%% Higgs field %%%%%%%%%%%%%%%%%%%%

The Higgs profiles $g(r)$ grow monotonically according to the boundary values (\ref{b1}) and (\ref{b2}), see the Fig. \ref{figg7}. The numerical analysis indicates that the core size inherent to these profiles is larger than those related to $G(\chi)\equiv 1$ for all values of $r_{0}$. Whether the value of the dielectric parameter increases within the interval $0\leq r_{0} \leq r^{\ast (2)}$, the core diminishes continuously until its minimum width attained for $r_{0}=r^{\ast(2)}$. In addition, when $r_{0}> r^{\ast(2)}$, the width of the core grows slowly. Here, for instance, whether one considers the width of the profile as $\approx 63.2\%$ of its height, the minimum width is attained for $r^{\ast(2)}\approx 1.1$.

%%%%%%%%%%%%%%%%%%% vector and magnetic field %%%%%%%%%%%%%%%%%%%%

Figure \ref{figg8} displays the solutions to both the gauge profile $a(r)$ and the magnetic field $B(r)$. In this case, the behavior of $a(r)$ on the parameter $r_{0}$ is analogous to previously described for the Higgs field. On the other hand, the magnetic field vanishing at the origin develops a global maximum, $B_{\text{max}}$, whose amplitude varies with $r_{0}$ as shown by the red line in the Fig. \ref{figg12}. It is worthwhile to note that for sufficiently large values of $r_{0}$, the amplitude $B_{\text{max}}$ decays as $r_{0}^{-1}$ and its localization, $r_{B_{\text{max}}}$, increases as $r_{0}^{1/2}$ along the radial coordinate, see the blue line in the Fig. \ref{figg12}.

%%%%%%%%%%%%%%%%%%% scalar potential and electric field %%%%%%%%%%%%%%%%%%%%

The results, which refer to the scalar potential $A_{0}(r)$ and the electric field, appear in Fig. \ref{figg9}. The amplitude $A_{0}(r=0)$ again depends on the dielectric parameter $r_{0}$, see the red line in the Fig. \ref{figg11}.  However, it now behaves in a different way than that related to the previous case; see Fig \ref{figg5}. The maximum amplitude of the electric field, $E_{\text{max}}$, versus the dielectric parameter $r_{0}$ is depicted in  the Fig. \ref{figg11}, while its radial localization, $r_{E_{\text{max}}}$,  which increases as $r_{0}^{1/2}$ for sufficiently large $r_{0}$, is shown as the black line in the Fig. \ref{figg12}.

%%%%%%%%%%%%%%%%%%% energy density $\varepsilon_{G}$ %%%%%%%%%%%%%%%%%%%%

The profiles of the energy density $\varepsilon _{G}(r)$ are plotted in Fig \ref{figg12}. The profile does not form a lump (as it happens when $G(\chi)\equiv 1$) but now develops a ring-like shape, such as it happens for the winding number $n\geq 2$ in the context of Maxwell-Higgs electrodynamics. Furthermore, the dependence of $\varepsilon_{G}(r=0)$ on $r_{0}$ is depicted as the black line in the Fig. \ref{figg11}, from which we see that it differs entirely from the one obtained in the first case, as shown by the Fig. \ref{figg5}.

%%%%%%%%%%%%%%%%%%%%%%%%

\section{Summary and perspectives}

We have investigated the formation of BPS solitons in a chiral medium ruled by an extended Maxwell-Chern-Simons electrodynamics coupled to the Higgs field and  an extra real scalar field $\chi $ which controls the dielectric medium.  In such a context, we have focused our attention on the BPS structure of the chiral model  in order to attain the self-dual BPS equations whose solutions saturate the Bogomol'nyi bound. In the sequence, we have observed that the self-dual equation for the  field $\chi$  does not involve the others sectors of the theory and depends solely on the particular choice for the
superpotential $F_{i}(\chi)$. Nevertheless,  the influence of the  field $\chi$ on the overall scenario occurs through the dielectric function $G(\chi)$, which appears in the equations for the gauge sector.

To describe the BPS vortices in the chiral medium, we have used the usual
radially symmetric ansatz for the Higgs and the gauge fields. Inspired by such maps, we have also parameterized the superpotentials $F_{i}(\chi)$ in terms of a function $W(r)$. We have solved the self-dual equation for $\chi(r)$ analytically and introduced an specific expression for the dielectric function $G(\chi)$ which becomes responsible for the changes on the profiles of the chiral vortices. To better understand the influence of the kink engendered by the field  $\chi$ through $G(\chi)$ on the vortex profiles, we have split our investigation into two cases based on the functional forms chosen for the dielectric function.

In all cases, the dielectric function satisfies the conditions $\sqrt{G}B\rightarrow 0$ and $\sqrt{G}\partial_i A_0 \rightarrow 0$ in the limit $|\mathbf{x}|\rightarrow \infty $,  which ensures the convergence of the total energy of the self-dual vortex to the Bogomol'nyi bound. The numerical analysis has revealed that the dielectric causes different effects on the corresponding vortex profiles in comparison to those obtained without it   (i.e., via $G=1$).  Our first choice, $G(\chi)=\chi^{-2}$, affects the behavior of the fields along the radial coordinate, except near the boundary values. In particular, the magnetic and the electric fields vanish at $r=r_{0}$,  from which the resulting configurations engender new ringlike profiles. Nevertheless, the tail of the chiral configurations decays like that of the Abrikosov-Nielsen-Olesen's vortex. On the other hand, the second choice $G(\chi )=(1-\chi ^{2})^{-1}$ changes the behavior of the fields near the boundaries. Consequently, the magnetic sector and the energy density develop ringlike formats. Nevertheless, the electric field preserves the same ring profile as in the canonical MCSH model. On the other hand, this time, the chiral vortex tail follows a power-law decay, which contrasts the exponential law inherent to the Abrikosov-Nielsen-Olesen's vortex.

An interesting perspective is the application of the present idea to the
study of the existence of chiral self-dual solitons in the context of $CP(2)$ or Skyrme models in the presence of a magnetic impurity, see the references \cite{15} and \cite{19}, for instance. We are investigating these
issues and will report the results in future contribution. 

\begin{acknowledgments}
This study was financed in part by the Coordena\c{c}\~{a}o de Aperfei\c{c}%
oamento de Pessoal de N\'{\i}vel Superior - Brasil (CAPES) - Finance Code
001, the Conselho Nacional de Desenvolvimento Cient\'{\i}fico e Tecnol\'{o}%
gico - Brasil (CNPq) and the Funda\c{c}\~{a}o de Amparo \`{a} Pesquisa e ao
Desenvolvimento Cient\'{\i}fico e Tecnol\'{o}gico do Maranh\~{a}o - Brasil
(FAPEMA). J. A. thanks the full support from CAPES. R. C. acknowledges the
support from the grants CNPq/306724/2019-7, FAPEMA/universal-00812/19, and FAPEMA/APP-12299/22. E. H. thanks the support from the grants CNPq/307545/2016-4 and FAPEMA/COOPI/07838/17.
\end{acknowledgments}


\begin{thebibliography}{99}
\bibitem{foot2} In what follows, we use $\eta ^{\mu\nu }=(+--)$ as the
metric signature of the $(1+2)$-dimensional flat spacetime, together with
the natural units system.
\end{thebibliography}
\end{document}